# Development of an Interactive Tutorial on Quantum Key Distribution

Seth DeVore and Chandralekha Singh

*Department of Physics and Astronomy, University of Pittsburgh, Pittsburgh, PA 15260*

**Abstract:** We describe the development of a Quantum Interactive Learning Tutorial (QuILT) on quantum key distribution, a context which involves a practical application of quantum mechanics. The QuILT helps upper-level undergraduate students learn quantum mechanics using a simple two state system and was developed based upon the findings of cognitive research and physics education research. One protocol used in the QuILT involves generating a random shared key over a public channel for encrypting and decrypting information using single photons with non-orthogonal polarization states, and another protocol makes use of two entangled spin-½ particles. The QuILT uses a guided approach and focuses on helping students build links between the formalism and conceptual aspects of quantum physics without compromising the technical content. We also discuss findings from a preliminary in-class evaluation.

**Keywords**: quantum mechanics, quantum key distribution, two state system, physics education research
**PACS**: 01.40Fk, 01.40.gb, 01.40G.

## INTRODUCTION

Quantum mechanics (QM) is a particularly challenging subject for undergraduate students. Based upon the research studies that have identified difficulties with quantum mechanics [1-4] and findings of cognitive research, we have developed a set of research-based learning tools to help students develop a good grasp of quantum mechanics. These learning tools include the Quantum Interactive Learning Tutorials (QuILTs) [5-6]. Here, we discuss the development and evaluation of a QuILT on quantum key distribution (QKD) [7].

QKD is an application of QM useful for generating a shared secure random key for encrypting and decrypting information over a public channel [7-8]. Secure QKD protocols involve two parties, a sender (Alice) and a receiver (Bob), who generate a random shared key over a public channel. A unique feature of these protocols is that Alice and Bob can detect the presence of an eavesdropper (Eve) who is attempting to gain access to the key by intercepting their communication during the shared key generation process.

Using a simple two state system, the QKD QuILT helps students learn foundational issues in QM including the fact that a quantum state can be in a superposition of linearly independent states and that measurement, in general, collapses the state into an eigenstate of an operator corresponding to the observable measured. The QKD QuILT provides a guided approach to help students learn about fundamental concepts of QM using protocols similar to those used by the banking industry [9] to allow secure data transfer over a public channel where someone could eavesdrop. The QuILT helps students learn that the ability to detect an eavesdropper in secure QKD protocols is due to the fact that physical observables are in general not well-defined in a given quantum state but measurement collapses the state and gives the observable a definite value. In particular, they learn that secure key generation exploits the fact that quantum measurement can disturb the state and that a random unknown quantum state cannot be cloned [7]. Students learn to reason that this indeterminacy is unique only to QM and can be used to determine if someone eavesdropped during the QKD process and if so, how much error would be introduced in the shared key generated by Alice and Bob due to the eavesdropper, Eve, even if she utilizes best practices. They learn that Eve must measure the state she intercepts and since quantum measurement can change the state, she will not always know what replacement quantum state to send to Bob each time she intercepts the communication between Alice and Bob. They learn that even if Eve uses clever strategies for sending replacement states, since she must often guess what replacement state to transmit to Bob, it will generate an error in the shared key that Alice and Bob generate in her presence and lead to Eve's presence being detected.

## THE DEVELOPMENT OF THE QUILT

The QuILT strives to help students learn QM using QKD with a simple two state system and was developed based upon the findings of cognitive research and physics education research. The development of the QKD QuILT began with an investigation of student difficulties with related concepts discussed later. The QuILT uses a guided approach in which various concepts build upon each other gradually. It strives to build connections between the formalism and conceptual aspects of QM without compromising technical issues. The development went through a cyclic interactive process which included the following stages: (1) development of the preliminary version based on a theoretical task analysis of the underlying knowledge structure and research on students' common difficulties with relevant concepts; (2) implementation and evaluation by administering it individually to students and by obtaining feedback from three faculty members who are experts in these topics; (3) determining its impact on student learning and assessing what difficulties were not adequately addressed in the preceding version; (4) making refinements based on feedback from the implementation.

Individual semi-structured interviews with fifteen undergraduate students enrolled in the QM course and graduate students were carried out using a think-aloud protocol to better understand the rationale for their responses throughout the development of various versions and extending to the pretest and posttest (which were given to students before and after they engaged in learning via the QuILT). In the QuILT, using a guided approach to learning, students are asked to predict what should happen in a particular situation, and after their prediction phase is complete, they are provided guidance. They are provided support to reconcile the differences between their predictions, and, e.g., what is provided in Fig. 1 for the B92 protocol [7]. In the second protocol for secure key generation using entangled particles, students can use a simulation to check their prediction. After each individual interview with a particular version of the QuILT, modifications were made based upon the feedback obtained from students. For example, if students got stuck at a particular point and could not make progress from one question to the next with the hints already provided, suitable modifications were made. We also iterated all components with three faculty members and made modifications based upon their feedback. When we found that the QKD QuILT was working well in individual administration and the posttest performance was significantly improved compared to the pretest performance, the first part of the QuILT that uses the B92 protocol with non-orthogonal polarization states

**TABLE 1.** An example of a table students must complete as part of the guided approach to learning in the QuILT

| Alice's polarization: | ↗ | ↔ | ↗ | ↔ | ↗ | ↔ |
|---|---|---|---|---|---|---|
| Bob's polarization: | ↙ | ↙ | ↕ | ↕ | ↕ | ↘ |
| Bob's detector clicks: | N | N | Y | N | N | Y |
| Bit is recorded: (Y or N) | | | | | | |
| Which bit they record: (0, 1 or -) | | | | | | |

B92 protocol with non-orthogonal polarization states of the photon was administered in class and the second part that uses entangled particles was given to students as homework after traditional instruction on relevant concepts including two state systems (e.g., spin-½, polarization states of photon) and addition of angular momentum (which was useful for understanding entangled states).

The QuILT first helps students learn the basics related to bits, qubits, polarization states of a photon, and effect of measurement on a two state system in a superposition of linearly independent states. The need for reinforcing these basics in a guided approach was evident during individual interviews with students. After working on the QuILT, one interviewed student stated "If I hadn't known the stuff in the basics I would be really confused with the QKD part".

Students are then guided through the B92 protocol [7] to generate a secure key in which Alice, Bob and Eve each encounter two non-orthogonal polarization states of a photon. Alice randomly sends single photons with one of the two non-orthogonal polarization states (e.g., with $0°$ and $-45°$ polarization states) and Bob randomly intercepts the single photons with one of two non-orthogonal polarization states (e.g., randomly with $90°$ and $-45°$ polarization states if Alice randomly transmits $0°$ and $-45°$ polarized photons) together with a 100% efficient photo-detector (detector) behind his polarizers to detect the photons transmitted by Alice. After the measurement of polarization, the photon is polarized in

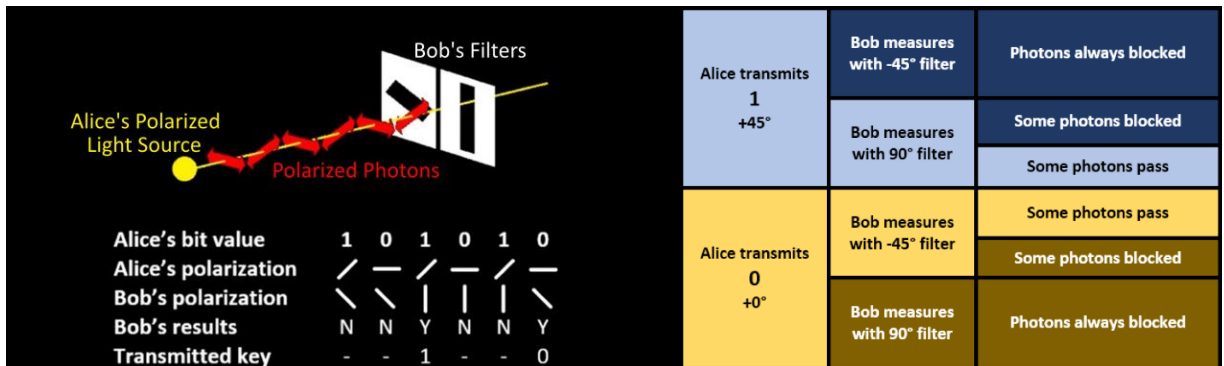
**FIGURE 1.** A figure in the latter part of the QuILT to help students check their predictions and reconcile the differences.

the state it was measured, with all information about its initial polarization lost. Students learn that a systematic comparison, e.g., of every 10$^{th}$ bit in the shared key, after a sufficiently long key is generated by each person will display at least a minimum threshold error if Eve was eavesdropping no matter how innovative her protocol for intercepting Alice's photons and replacing them is. In the QuILT, students are first guided through a series of questions in which there is no eavesdropper. These questions culminate in Table 1 which students must construct that spans all possible combinations of Alice's polarization, Bob's polarization, whether or not Bob's detector clicks and whether a bit is generated. Students are then guided to address cases in which Eve has eavesdropped on the key generation process. The questions lead students to reason about how Eve's quantum measurement introduces error in the shared key generated and how much error is generated. The QKD QuILT Part II which helps students learn the BBM92 [8] protocol involving entangled states of two spin-½ particles for secure key generation will not be discussed here. Below, we discuss common student difficulties addressed in the QuILT.

## STUDENT DIFFICULTIES

**Difficulty with polarization states of photons:** During the interviews throughout the development of the QuILT, some students claimed that the polarization states of a photon cannot be used as basis vectors for a two state system due to the fact that a photon can have any polarization state. They argued that since a polarizer can have any orientation and the orientation of the polarizer determines the polarization state of a photon after it passes through the polarizer, it did not make sense to think about polarization states of a photon as a two state system. These students were often so fixated on their experiences with polarizers from everyday life (which can be rotated to make their polarization axis whichever way one wants with respect to the direction of propagation and polarization of incident light) that they had difficulty thinking about polarization states of a photon as vectors in a two-dimensional space.

It is interesting to note that most students who had difficulty accepting that the polarization states of a photon can be used as basis states for a two state system had no difficulty accepting that spin states of a spin-½ particle can be used as basis states for a two state system despite the fact that these two systems are analogous from an expert perspective. Interviews suggest that this difference in their perception was often due to how a spin-½ system and polarization were first introduced and the kinds of mental models students had built about each system. Generally, students are introduced to polarization in an introductory course and to spin-½ systems in a QM course. Discussions suggest that some students were so used to the classical ways of thinking about a beam of light passing through a polarizer that they had difficulty thinking about the polarization states of a photon as vectors in a two dimensional Hilbert space. Since students had learned about the spin-½ system only in QM, thinking of spin states of a spin-½ particle as vectors in a two dimensional space often did not create a similar conflict.

**Difficulty with the probabilistic nature of interactions between a single photon and a polarizer:** Some interviewed students also incorrectly claimed that whenever single photons with a given polarization were sent through a polarizer, they would be completely blocked (absorbed) by the polarizer only when the photon polarization was orthogonal to the polarization axis. They therefore incorrectly claimed that a photon with any other polarization would always make the detector behind the polarizer click. Individual discussions suggest that this difficulty was often related to difficulty in applying the measurement postulate of QM to the situation of a single photon incident on a polarizer. Students sometimes claimed that a single photon incident on a polarizer can partly pass through the polarizer and partly get absorbed. They did not realize that a single photon will either get absorbed by the polarizer or it will pass through. Interviews suggest that this difficulty often originated from the fact that students were not considering single photons passing through a polarizer probabilistically and were interpreting the situation by mixing quantum mechanical and classical ideas. Discussions during the interviews also suggest that sometimes student difficulty even in this case was coupled to the confusion between a single photon and a beam of light incident on a polarizer. In particular, students were overgeneralizing their knowledge that the intensity of light generally decreases after passing through a polarizer and it is only when the polarization of incident light is orthogonal to the polarization axis of the polarizer that the light is completely absorbed.

**Assuming that Bob always knows the photon polarization:** Related to the difficulties with the photon polarization states discussed above, a common difficulty before working on the QKD QuILT was believing that Bob always knows the polarization of the photon sent by Alice in the B92 protocol. Some students incorrectly claimed that it is possible for Bob to determine the polarization state of Alice's photon both when a photon is absorbed by Bob's polarizer (not detected by the photo-detector behind Bob's polarizer) or it is transmitted by Bob's polarizer (detected by the detector behind Bob's polarizer). Students who had this difficulty incorrectly claimed that Bob will know the polarization of the photon even if his detector does not click due to their incorrect assumption that the photon polarization must be perpendicular to the polarization axis of the polarizer for the photon to be completely absorbed. These students incorrectly stated that whether the detector behind Bob's polarizer clicks or not, Bob knows the polarization of the photon sent by Alice. They struggled with the fact that for each photon polarization that Alice sends, there is a non-zero probability of the photon being absorbed completely by Bob's polarizer due to quantum measurement collapsing the polarization state. Similarly they had difficulty recognizing that only one of the two possible polarizations sent by Alice has a probability of being transmitted through each of Bob's two polarizers that he randomly uses for making his measurements of the photon polarization.

**Assuming that Bob knows the photon polarization only when his detector does not click:** Other students incorrectly claimed that Bob is only certain about the polarization when the photon sent by Alice is polarized perpendicular to the polarization axis of his polarizer. They struggled with the fact that Bob will know about Alice's photon polarization only when the photon passes through his polarizer and his photo-detector registers the photon. They incorrectly claimed that since a photon with a polarization perpendicular to the polarization axis always results in the photon being blocked, Bob must always know the polarization of the photon Alice sent in this case when the photon is always blocked. In the case in which a photon's polarization is neither perpendicular nor parallel to Bob's polarizer's axis, they incorrectly claimed that since there is a non-zero likelihood for the photon to be both transmitted and absorbed or to be either transmitted or absorbed, Bob cannot be sure of the photon polarization sent by Alice.

## RESULTS FROM PRE/POSTTEST

Once we determined that the QuILT was effective in individual administration, it was administered in three upper-level QM courses. Students were given a full quiz score for trying their best on the pretest administered in class to 49 students before they worked on the QuILT. Before the pretest, students had learned about spin-½ systems, addition of angular momentum, polarization states of photons and a basic outline of quantum key distribution. They then worked through the QuILT in class in small groups. The posttest, which was counted as a quiz grade, was administered in the next class to 50 students after all students had the opportunity to complete the QuILT at home if they did not finish it in class. The pretest and posttest are identical in structure with only the polarization angles changed. Each begins by outlining a situation similar to the B92 protocol to ensure that the students are familiar with it. Each asks a series of questions about quantum measurements regarding a B92 protocol setup [7] with Alice's polarization axes randomly switched between 70° or 0° on the pretest (60° or 0° on the posttest) and Bob's polarizer axes randomly switched between -20° or 90° on the pretest (-30° or 90° on the posttest).

The average scores on the pre/posttests were 56% and 91%, respectively (49 and 50 students). Questions 1 and 2 on the pretest and posttest are great indicators of the effectiveness of the QuILT as they demonstrate the two possible outcomes of an individual measurement made by Bob. Question 1 provides the angle of Bob's polarizer, informs students that Bob's photo-detector does not click and asks them to identify what Bob knows about the polarization of the photon sent by Alice in this situation. On the pretest, 41% of students incorrectly claimed that Bob will know the polarization of the photon sent by Alice and a "bit" of the key can be generated even when a photon is not detected by Bob's detector but on the posttest only 6% had this difficulty. Question 2 is similar to Question 1 except that Bob's photo-detector clicks. When the detector clicks, Bob knows that the photon that Alice sent cannot be polarized perpendicular to Bob's polarizer and therefore he can determine the polarization of the photon Alice sent. On the pretest and posttest, 37% and 6% of the students struggled on this question, respectively. Student performance on other questions also improved significantly from the pretest to posttest.

## SUMMARY

We developed a research-based QuILT to help upper-level undergraduate students learn QM in the context of QKD. The QuILT uses a guided approach to bridge the gap between the quantitative and conceptual aspect of QM. It strives to help students build a good knowledge structure of foundational issues in QM using the context of quantum protocols for secure key generation for which there is no classical analog. The preliminary evaluation shows that the QuILT helps improve student understanding of related concepts.

## ACKNOWLEDGEMENTS

We thank the National Science Foundation for awards PHY-0968891 and PHY-1202909.